# Geometric resonance in intermediate state of type I superconductors

Oleg P. Ledenyov

*National Scientific Centre Kharkov Institute of Physics and Technology,
Academicheskaya 1, Kharkov 61108, Ukraine.*

The attenuation of ultrasound (the geometric resonance) in an intermediate state of the *type I* superconductors at $ka_N >> 1$ is researched. The oscillating dependence of ultrasound attenuation, which has an amplitude modulation as a result of presence of the Andreev reflections by the electronic excitations on the boundaries between the normal metal layer and the superconducting layer in an intermediate state of the *type I* superconductor, is obtained. The derived theoretical equation explains the nature of experimental results in [1, 2].

## Introduction

The research interest to the study on the ultrasonic signal attenuation in an intermediate state of *type I* superconductors is stipulated by the fact that this is the only method, which allows to obtain the experimental information on the internal structure of the intermediate state of *type I* superconductors at external magnetic fields. The electronic excitations in the normal metal layers mainly interact with the ultrasonic wave in an intermediate state of *type I* superconductors. These electronic excitations have a distinctive feature, which is represented by a special mechanism of the electronic excitations reflection on the normal metal – superconductor boundaries in an intermediate state of *type I* superconductors as it was shown by Andreev in [3]. Going from the Andreev dynamics of electronic excitations in an intermediate state of *type I* superconductors, the theoretical researches on the propagation of ultrasonic wave in an intermediate state of *type I* superconductors were completed [4-9]. In [4], the mechanism of ultrasound attenuation, connected with the vibration of the boundaries between the phases, was discovered. This mechanism was subsequently applied to research the case of the threads like structure in an intermediate state of *type I* superconductors [5]. In 1967, Andreev first forecasted the phenomena of oscillating attenuation of ultrasonic wave [6], and made the appropriate calculations by solving the kinetic equation with the application of the Fourier transform components, using the inverse value with the double thickness of normal metal layer. At $ka_N \sim 1$ and $l, D > a_N$, the periodicity of ultrasound attenuation in the dependence on $a_N$ was found, where $a_N$ is the thickness of normal metal layer, $l$ is the electron mean free path and $D$ is the diameter of electronic excitations orbit in the critical magnetic field $H_c$. The influence by the magnetic quantization on the ultrasound attenuation [7, 8] and the features of ultrasonic waves propagation in the case of their tilted falling on the intermediate state structure in *type I* superconductors [9] were also considered.

## Geometric resonance in intermediate state of type I superconductors

The propagation of longitudinal ultrasonic wave in an intermediate state of *type I* superconductors is theoretically analyzed in the frames of a simple model of the closed Fermi surface with the two points of rotation on the extreme cross-section, assuming that the following expressions are true $l > D > a_N >> \lambda$, where $\lambda$ is the wavelength of ultrasonic signal. The ultrasonic wave with the wave vector $\mathbf{k}$ propagates across the system of alternating normal metal and superconducting layers toward the direction of the axe *X*. The critical magnetic field $\mathbf{H}_c$ is oriented toward the direction of the axe *Z*. The separation boundary between the normal metal phase and the superconducting phase (*N-S* boundary) lies in the plane *YZ*.

The kinetic equation in the field of deformation of ultrasonic wave $u_{ik} = u_{ik,0} \exp(i\mathbf{k}\mathbf{r} - i\omega t)$ can be written in the normal metal layer as [10, 11]

$$(i\mathbf{k}\mathbf{v} - i\omega + \nu)\psi + \frac{\partial \psi}{\partial t_1} = \Lambda_{ik} \dot{u}_{ik} \qquad (1)$$

where $\Psi$ can be found from $f = f_0 - \frac{\partial f_0}{\partial \varepsilon}\psi$ ($f_0$ and $f$ are the equilibrium and nonequilibrium functions of the electrons distribution) [11]. In eq. *(1)*, in the view of the fact that the sound velocity is small in comparison with the electron velocity **v**, let us disregard the time derivative of $\Psi$, which is proportional to the $\omega$. Let us search for the solution of the eq. *(1)* in the normal metal layer as shown in eq. *(2)*

$$\psi(t_1) = \int_{-\infty}^{t_1} \Lambda_{ik}(t_2)\dot{u}_{ik}(t_2) \exp\left\{\int_{t_1}^{t_2}(i\mathbf{k}\mathbf{v}(t_3) + \nu)dt_3\right\} dt_2 \quad (2)$$

Let us note that, in the system of coordinates, connected with the electronic excitation $\dot{u}_{ik} \sim (\mathbf{kv} - \omega)$ at $\mathbf{kv} \gg \omega$, the expression $\psi(v) + \psi(-v) = 0$ is true at the change of sign $\mathbf{v} \to -\mathbf{v}$ on the N-S boundary, which is required to satisfy the Andreev boundary conditions.

Taking to the account the relation $\mathbf{k} \perp \mathbf{H}$ and the presence of the periodicity in the selected geometry $\Lambda_{ik}\dot{u}_{ik}$ in the relation to the period of rotation of electronic excitation in an intermediate state of type I superconductor in magnetic field, the expression for the coefficient of ultrasound attenuation in an intermediate state of type I superconductor in magnetic field, which is formally analogous to the expression for the coefficient of ultrasound attenuation in normal metal, can written as

$$\Gamma(X_1) = \frac{2}{(2\pi\hbar)^3} A \int \frac{\Omega}{\mu} \left\{ |\gamma_{(1)}|^2 + |\gamma_{(2)}|^2 + \gamma_{(1)}^* \gamma_{(2)} \sin\left(\int_{t_{(1)}}^{t_{(2)}} \mathbf{kv} dt\right) \right\} dp_z \quad (3)$$

where $A$ is the coefficient, which depends on the physical properties of the propagation medium and the frequency of ultrasonic signal [10, 11], $\gamma_{(l)} = \Lambda_{ik(l)} \dfrac{\dot{u}_{ik}}{|\mathbf{kv}'_{(l)}|^{1/2}}$ , the lower index in the brackets provides an information on the selected turn point from the two considered turn points in which $\mathbf{kv}_{(1),(2)}=0$, where the magnitudes of corresponding parameters have to be taken for the calculation. In an intermediate state of the type I superconductors, the ultrasound attenuation coefficient $\Gamma(X_{(1)})$ occurs to be dependent on the position at the electronic excitation's orbit at the axe X, that is on the coordinates of the turn point $X_1$.

Let us divide the integral under the sign of sinus in the expression for an oscillating part of ultrasound signal attenuation on a sum of integrals with the limits set at the subsequently fixed times of collisions by an electronic excitation with the N-S boundaries as

$$\int_{t_{(1)}}^{t_{(2)}} \ldots = \int_{t_{(1)}}^{\tau_1} \ldots + \int_{\tau_1}^{\tau_2} \ldots + \ldots + \int_{\tau_n}^{t_{(2)}} \ldots .$$

Taking to the consideration that the integral $\int_{t_{(1)}}^{t_{(2)}} |\mathbf{kv}| dt = kD$, let us calculate the integral $\int_{t_{(1)}}^{t_{(2)}} |\mathbf{kv}| dt = kd$ , where $d$ is the certain effective diameter of the trajectory by an electronic excitation, introduced in an analogy with the normal metal as shown in Fig. 1.

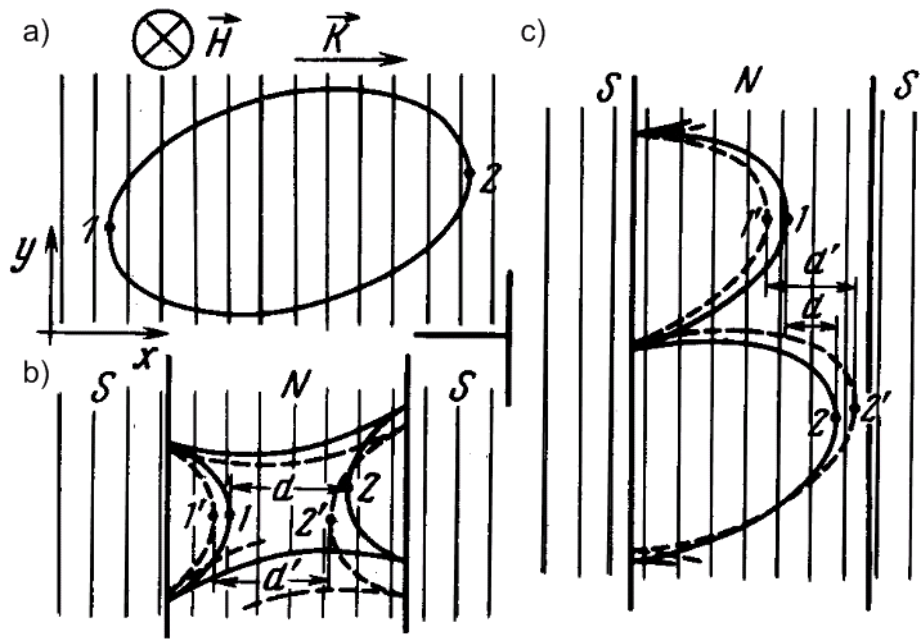

*Fig. 1. Trajectory by electronic excitation in type I superconductor in magnetic field:*
*a) Trajectory by electronic excitation in normal metal layer in type I superconductor in magnetic field;*
*b) Trajectories by electronic excitation in an intermediate state in type I superconductor in magnetic field, when there are two reflections by electronic excitation between turn points 1 and 2. Distance between 1 and 2 along axe X is equal to d and it does not depend on position of point 1;*
*c) Trajectories by electronic excitation in an intermediate state in type I superconductor in magnetic field, when there is one reflection by electronic excitation between turn points of 1 and 2. Distance between point 1 and point 2 along axe X is equal to d and it depends on position of point 1, $d \neq d'$.*
Thin vertical lines denote planes with equal phases of ultrasonic wave.

In the case of an even number of the reflections by an electronic excitation on the *N-S* boundaries between the turn points, the certain effective diameter *d* of trajectory by an electronic excitation is equal to

$$d = (-1)^l \{(l+1)2a_N - D\}, \quad (4)$$

where $l = \left[\dfrac{D-a_N}{2a_N}\right]$, and [...] is the whole part of the expression in the brackets (*[x] = –1*, if *–1 ≤ x < 0*). The graph with the dependence, characterizing the certain effective diameter *d* of the trajectory by an electronic excitation as a function of *D/2a_N* such as $d = f\left(\dfrac{D}{2a_N}\right)$, which is shown in Fig. 2.

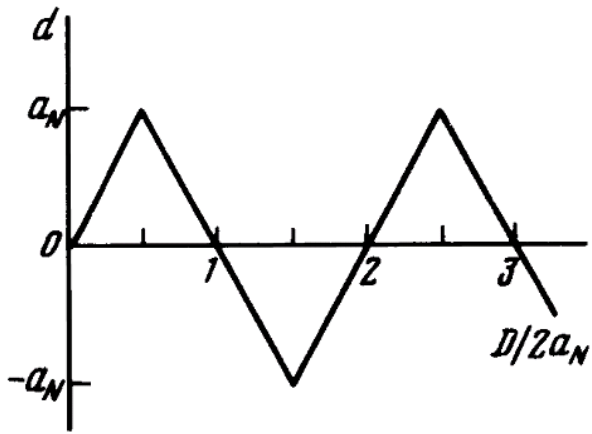

*Fig. 2. Dependence of distance d between turn points along axe X at even number of reflections by electronic excitations on relation D/2a_N.*

In the case of an uneven number of the reflections by an electronic excitation on the *N-S* boundaries between the turn points, the certain effective diameter *d* of the trajectory by an electronic excitation will be represented as a partly linear function of $X_1$ with the measurements limits from $-a_N$ up to $a_N$, and the sort of this dependence is not significant in this case.

The calculation of the average coefficient of ultrasound attenuation $\langle\tilde{\Gamma}\rangle$ can be done by the averaging of an oscillating part of ultrasound attenuation $\tilde{\Gamma}(X_{(1)})$ at $ka_N \gg 1$ as in the following expression

$$\langle\tilde{\Gamma}\rangle = \dfrac{\int_0^{a_N} \tilde{\Gamma}(X_{(1)}) dX_{(1)}}{\int_0^{a_N} dX_{(1)}}.$$

In view of the fact that a number of reflections by an electronic excitation is a function of $X_{(1)}$ at the given $a_N$ and *D*, therefore the integral in the numerator decomposes into the two integrals, including the orbits with an even numbers of reflections and an odd number of reflections. The coefficient of ultrasound attenuation $\tilde{\Gamma}(even)$ does not depend on the $X_{(1)}$, and the problem reduces to the calculation of the integral $\int_0^{a_N} dX_{(1)}(even)$, which is equal to $a_N - |d|$. The coefficient of ultrasound attenuation $\tilde{\Gamma}(odd)$ is a fast oscillating function of the $X_{(1)}$ with the alternating sign, which converges to the zero. Thus, the average coefficient of ultrasound attenuation can be described by the modulated function $\langle\tilde{\Gamma}\rangle = \dfrac{a_N - |d|}{a_N}\hat{\Gamma}(even)$. Let us write the expression for the $|d|$

$$|d| = (-1)^n \left(D - \left\{m + \dfrac{1}{2}(1-(-1)^n)\right\}2a_N\right), \quad (5)$$

where $m = \left[\dfrac{D}{2a_N}\right]$, $n = \left[\dfrac{D}{a_N}\right]$.

The oscillating part of the coefficient of ultrasound signal attenuation in an intermediate state of *type I* superconductor is

$$\tilde{\Gamma} \sim \Delta\tilde{\Gamma}_{norm.\ metal}(H_C)\eta\dfrac{a_N - |d|}{a_N}\sin\left(kd \pm \dfrac{\pi}{4}\right) \quad (6)$$

where $\Delta\tilde{\Gamma}_{norm.\ metal}(H_C)$ is the amplitude of the oscillations of the geometrical resonance in the normal metal at the critical magnetic field $H_c$, $\eta$ is the concentration of the normal metal phase. It is visible in eq. (6) that the number of oscillations of ultrasound attenuation at the one period of modulation is equal to the double thickness of normal metal layer $2a_N$, expressed in the units of ultrasonic signal wavelength, that is $2a_N = n\lambda$, where *n* is the number of oscillations during the one period of modulated ultrasonic signal. In the case, when $D \ll a_N$, the results can be written as in the expression for the normal metal $d \to D$, $(a_N - |d|)/a_N \to 1$, and the concentration dependence of amplitude of oscillations remains in eq. (6). This concentration dependence of amplitude of oscillations of ultrasound attenuation can be used for the precise evaluation of the concentration of the normal metal phase in an intermediate state of the *type I* superconductors.

## Conclusion

The theoretical research on the attenuation of ultrasound (the geometric resonance) in an intermediate state of the *type I* superconductors at $ka_N \gg 1$ is completed. The equation to describe the oscillating dependence of ultrasound attenuation, which is modulated by the amplitude, because of the presence of the Andreev reflections by the electronic excitations on the normal metal – superconductor (*N-S*) boundaries in an intermediate state of the *type I* superconductor, is derived. The completed comparison of the theoretical results with the experimental data shows that the obtained formula (6) perfectly describes the complex periodicity of experimentally observed dependences [1, 2]. Let note that, in an agreement with the formula (6), the thickness of a layer with the normal metal phase changes from $10^{-2} cm$ to $10^{-3} cm$, depending on the magnitude of applied external magnetic field in an intermediate state of the *type I* superconductor. The detailed quantitative comparison of the theoretical results with the experimental outcomes [1, 2] will be conducted in the next research paper.

Author thanks to B. G. Lazarev for the research problem formulation and his kind support, A. F. Andreev for the thoughtful discussion on the research results, G. D. Filimonov and V. A. Shklovsky for their strong interest in the innovative research.

This research paper was published in the *Letters to the Journal of Experimental and Theoretical Physics* (*JETP Letters*) in 1979 [12].

*E-mails: ledenyov@kipt.kharkov.ua

_______________

1. A. G. Shepelev, O. P. Ledenyov, G. D. Filimonov, New Effects in Attenuation of Ultrasound in Intermediate State of High Pure Superconductor, *JETP Letters*, v. **14**, p. 428, 1971.
2. A. G. Shepelev, O. P. Ledenyov, G. D. Filimonov, Experimental Research on Longitudinal Ultrasound Attenuation in Intermediate State of High Pure Type I Superconductor, *Problems of Atomic Science and Technology* (*VANT*), Series «*Fundamental and Applied Superconductivity*», №1, p. 3, 1973.
3. A. F. Andreev, *JETP*, v. **46**, p. 1823, 1964.
4. A. F. Andreev, Yu. M. Bruk, *JETP*, v. **50**, p. 1420, 1966.
5. Yu. M. Bruk, *Journal of Low Temperature Physics* (*FNT*), v. **2**, p. 1130, 1976.
6. A. F. Andreev, *JETP*, v. **53**, p. 680, 1967.
7. V. P. Galaiko, E. V. Bezuglyi, *JETP*, v. **60**, p. 1471, 1971.
8. G. A. Gogadze, I. O. Kulik, *JETP*, v. **60**, p. 1819, 1971.
9. A. G. Aronov, A. S. Ioselevich, *JETP*, v. **74**, p. 580, 1978.
10. V. L. Gurevich, *JETP*, v. **37**, p. 71, 1959.
11. A. A. Abricosov, Introduction to the Theory of Normal Metals, *Nauka*, Moscow, Russian Federation, 1972.
12. O. P. Ledenyov, Geometric Resonance in Intermediate State of Superconductors, *JETP Letters*, v. **30**, issue 3, pp. 185-189, 1979 (in Russian).
www.jetpletters.ac.ru/ps/434/article_6857.pdf